\documentstyle[12pt]{article}

\advance \textheight by \topskip
\title{Twisted Family Structure and Neutrino Large Mixing}

\author{Masako BANDO, Physics Division, Aichi University, \\
Miyoshi, Aichi, Japan, 470-0296}

\begin{document}
\maketitle

\begin{abstract}
I demonstrate that neutrino large mixing between $\nu_\mu$ and
$\nu_\tau$ are naturally reproduced using a novel mechanism
called `E-twisting'in a supersymmetric $E_6$ grand unification
model. This model explains  all the characteristic features of
the quark/lepton Dirac masses as well as the neutrino's Majorana
masses despite the fact that all the members in {\bf27} of each
generation are assigned a common family charge. Most remarkably,
this  model yields a novel relation which gives the 2-3 lepton
mixing angle $\theta_{\mu\tau}$ in terms of quark masses and CKM
mixing: $\tan\theta_{\mu\tau}=(m_b/m_s)V_{cb}$, which is a kind
of $SO(10)$ GUT relation similar to the celebrated $SU(5)$
bottom-tau mass ratio. This relation is a result of a common
`twisted $SO(10)$' structure \footnote{Talk given at the
Internatinal Workshop on Neutrino Oscillation and their Origin, 
Fujiyoshida, Japan, February 11-13, 2000}. 
\end{abstract}
A remarkable fact  observed in SuperKamiokande~\cite{SK} 
is the very large lepton mixing, which is 
in a sharp contrast to the quark sector where the CKM
mixings are all small. Why can  such a large difference 
occurs between the quark and lepton sectors? 
This is  a challenging question for any particle physicist 
who tries to find grand unified theories (GUTs). 
Clearly any GUT which treats the
three families of quarks and leptons as a mere repetition no longer
works. We need some new mechanism of family structure. 
Lots of proposals  have been made on 
the origin of this large lepton mixing angles~\cite{model}.
On the other hand, the SK results indicate
larger unification groups than $SU(5)$ 
including left-right symmetry, in which large neutrino mixings seem 
unnatural, since in such larger GUT groups, for example, 
 $SO(10)$ GUT, all the fermions of a family are combined into 
a single representation, and the most natural prediction would be that the 
neutrino mixing is also very small with hierarchical masses.  

Recently we have constructed a supersymmetric $E_6$ unified model 
with an extra $U(1)$ family charge\cite{BKY}. There we have shown 
that E-twisted  family structure can reproduce all the characteristic features 
of the fermion masses, not only the quark/lepton Dirac masses but also the
neutrino Majorana masses. Despite the fact that a common $U(1)$ charge
is assigned to all the members in a fundamental representation {\bf 27} of 
$E_6$ for each family, the model well explains all the qualitative 
feature of different mass hierarchies among families and  between up and 
down quark sectors, as well as the mixing angles.
In this scenario we have found that in  the framework of a supersymmetric 
$E_6$ grand unified model~\cite{BKY}, the twisted family structure
yields a novel relation
\begin{equation}
  \tan\theta_{\mu\tau}\;=\;{m_b\over m_s}V_{cb},  
  \label{novel}
\end{equation}
which  I shall compare with the experiments later. 
Leaving the details in the papers~\cite{BKY}, I here explain the 
essence of the model. 
In the supersymmetric $E_6$ model, 
in addition to $E_6$ gauge vector multiplet,
we introduce chiral matter multiplets corresponding to the three families,  
 ($\Psi_i$ $(i=1,2,3)$) and a pair of Higgs fields, which is introduced mainly 
 for the electroweak symmetry breaking,($H$, $\bar H$)
\footnote{In order to give all the unwanted fermions to get heavy masses. 
we need another Higgs pair,  ($\Phi$, $\bar\Phi$) and which are 
responsible for realizing the E-twisted family  structure. 
Also we have to add 
a chiral Higgs multiplet $\phi({\bf 78})$ in order to 
break the GUT to the standard gauge group. 
Here we neglect those fields and start with the low energy fermions 
realized in twisted family structure.}.
In table 1, we summarize
all the fields we need in our model.
\begin{table}[htbp]
\begin{center}
\begin{tabular}{c|c|c|c|c|c|c|c|c|c}\hline
  & $\Psi_1$ & $\Psi_2$ & $\Psi_3$ & $H$ & $\bar H$ & $\Phi$ &
  $\bar\Phi$ & $\phi$ & $\Theta$ \\ \hline 
  $E_6$ & {\bf 27} & {\bf 27} & {\bf 27} & {\bf 27} 
  & ${\bf 27^*}$ & {\bf 27} & ${\bf 27^*}$ & {\bf 78} & {\bf 1} \\
  \hline
  $U(1)$ charge & 3 & 2 & 0 & 0 & 0 & $-4$ & 4 & $-2$ & $-1$ \\
  \hline
  $R$ parity & $-$ & $-$ & $-$ & + & + & + & + & + & + \\ \hline
\end{tabular}
\caption{$E_6$ representations and $U(1)$ charge assignment.}
\end{center}
\end{table}
The $E_6$ singlet field $\Theta$ with $U(1)$ charge $-1$ plays an
important role that its suitable powers compensate the mismatch of 
the $U(1)$ charge in the superpotential interaction terms. The $U(1)$
flavor symmetry discriminates families  and induces hierarchy
between them. Note that all the quarks and leptons in
one generation have a common $U(1)$ quantum number.

The following Yukawa superpotentials which are invariant 
under $R$ parity, $U(1)$ and $E_6$
will give masses of matter superfields $\Psi_i({\bf 27})$
\footnote{
There are other superpotentials including the Higgs fields, 
$\phi({\bf 78})$ and $\Phi({\bf 27})$, whose family charges are  not 
zero and will contribute to the  Yukawa terms of  the 2nd and 1st families.},
\begin{equation}
  W_Y(H) = y_{ij}\,\Psi_i({\bf 27})\Psi_j({\bf 27})H({\bf 27})
  \left({\Theta\over M_P}\right)^{f_i+f_j},
  \label{yukawa}
\end{equation}
where $f_i$ denotes the $U(1)$  charge of $i$-th family. With the 
coupling constants $y$  of order 1, 
the effective Yukawa coupling constants are associated with 
additional powers of $\lambda=\langle\Theta\rangle/M_P$\cite{FN}, 
which we assume 
is  of the order of the Cabibbo angle $\lambda\sim0.22$.
 We also suppose that only 
the $SU(2)$ doublet components of $H$ can have the electroweak scale
vacuum expectation value (VEV).

An interesting fact is that there are two ${\bf 5}^*$'s of $SU(5)$ in 
each ${\bf 27}$, i.e., ${\bf 5}^*$ of ${\bf 16}$ of $SO(10)$ ((${\bf 16,5^*}$))
and ${\bf 5}^*$ of ${\bf 10}$ ((${\bf 10,5^*}$)). Those may be called 
`E-parity' doublet. It is this doubling that we have a freedom to
choose the low-energy ${\bf 5}^*$ candidates. This actually implies
that the embedding of $SO(10)$ into $E_6$ such 
that $SU(5)_{\rm GG}\subset SO(10)\subset E_6$ with 
Georgi-Glashow $SU(5)_{\rm GG}$, possesses a freedom of rotation 
of $SU(2)_R$. The doubling of ${\bf 5}^*$'s in each ${\bf 27}$ also
provides the low-energy surviving down-type Higgs field with the
freedom of mixing parameter between two ${\bf 5}^*$'s 
in $H({\bf 27})$:
\begin{eqnarray}
  H({\bf 5}^*) &=& H({\bf 10,5^*})\cos\theta+H({\bf 16,5^*})
  \sin\theta.
 \label{higgsmixing}
\end{eqnarray}

Now we pick up the low-energy matter fields among the 
three $\Psi_i({\bf 27})$ of the above. The up-quark sector is unique
since ${\bf 10}$ and ${\bf 5}$ of $SU(5)$ appear only once in 
each ${\bf 27}$. As for the down
quarks, there is a freedom for choosing three from six ${\bf 5}^*$s in
three $\Psi_i({\bf 27})$s. We have classified possible typical
scenarios in Ref.~\cite{BKY}; (i) Parallel family structure, (ii)
Non-parallel family structure, and (iii) E-twisted structure. Among
these three possibilities, we here take the simplest and most
attractive option, namely the E-twisted family:
\begin{eqnarray}
  ({\bf 5}^*_1,\ {\bf 5}^*_2,\ {\bf 5}^*_3) &=& 
  \bigl(\Psi_1({\bf 16,5^*}),\ \Psi_2({\bf 16,5^*}),\ 
  \Psi_3({\bf 10,5^*})\bigr).
  \label{twist}
\end{eqnarray}
This structure implies that the third family ${\bf 5}^*$ belongs to 
 ${\bf 10}$ of $SO(10)$. This twisting is realized by the suitable
 VEVs of the Higgs fields( the details will be found in Ref.~\cite{BKY}).

Let us here concentrate ourselves on the 2nd and 3rd families and see
what happens to their  masses and mixings. 
The $2\times2$ mass matirces  for the up-quark, down-quark and 
charged-lepton, $M_u$, $M_d$ 
and $M_e$, are expressed as,  
\begin{equation}
  M_u = \bordermatrix{
        & u^c_2  & u^c_3 \cr
    u_2 & y_{22} & y_{23} \cr
    u_3 & y_{32} &  y_{33}\cr}
                   v \sin\beta
= \bordermatrix{
        & u^c_2  & u^c_3 \cr
    u_2 & *      &  f\lambda^2 \cr
    u_3 & *      &  1\cr}
                   vy_{33} \sin\beta, 
  \label{mu}
\end{equation}

\begin{equation}
  M_d = \bordermatrix{
      & d'^c_2                  & D^c_3 \cr
 d_2  & z^d_{22}\cos\theta      &y_{23} \sin\theta\cr
 d_3  & z_{32}\cos\theta      & y_{33} \sin\theta\cr}  
 v\cos\beta
= \bordermatrix{
     & d'^c_2                  & D^c_3 \cr
 d_2  & e\lambda^2      &f\lambda^2\cr
 d_3  & h      & 1\cr}  
 v y_{33} \sin\theta\cos\beta.
  \label{md}
\end{equation}

\begin{equation}
  M_e^{\rm T} =
  \bordermatrix{
        & e'_2 & E_3 \cr
  e^c_2 & z^e_{22}\cos\theta   &y^e_{23} \sin\theta\cr
  e^c_3 & z_{32}q\cos\theta& y_{33}  \sin\theta\cr} 
                          v \cos\beta
=\bordermatrix{
        & e'_2 & E_3 \cr
  e^c_2 & *        & *       \cr
  e^c_3 & h        & 1    \cr} 
                          v y_{33}  \sin\theta \cos\beta. 
  \label{me}
\end{equation}
where $\tan\beta$ is the mixing angle of two light Higgs doublets 
and $v$ is the VEV of the standard Higgs
field $H( {\bf 27})$. We rewrite by using simple notations 
in the third terms  with  $*$ being irrelevant for our present 
discussions
\footnote{We have assumed that the main contribution comes only from 
the Higgs field  $H( {\bf 27})$ at least for the quark mass matrix 
of $33$ and $23$ elements in the quark mass matrices, $ M_u$ and 
$M_d$.}.
Note that by taking the mixing $\sin\theta$ 
of  the Higgs field $H( {\bf 27})$ of Eq.(\ref{higgsmixing}), 
to be of order $\lambda^2$, $h$ becomes of order $1$. 
It is easy to obtain  $h$ from the bottom and strange quark masses and 
mixing angle, $m_b, m_s, V_{cb}$ from Eqs.(\ref{mu}) and ((\ref{md})). 
Noting that 
$h$ gives directly the lepton mixing angle 
$\tan \theta _{\mu \tau}=h$
\footnote{We can confirm that the right handed Majorana mass term 
indicates very small mixing and neutrino mixing mainly comes from
 lepton mixing.},
we can get the novel relation Eq.(\ref{novel}),  or equivalently,
\begin{eqnarray}
  \sin^2 2\theta_{\mu\tau} &=& \frac{4V_{cb}^2
    \left(\displaystyle\frac{m_s}{m_b}\right)^2}{\left[V_{cb}^2 
      +\left(\displaystyle\frac{m_s}{m_b}\right)^2\right]^2}\,.
  \label{sumrule}
\end{eqnarray}
By taking  the experimental 
value of  $x=V_{cb}m_b/m_S$, $1\leq  x\leq 1.68$, 
we can calculate the left hand side of Eq.(\ref{sumrule}), namely,  
\begin{equation}
0.78\leq  \sin^2 2\theta_{\mu\tau} \leq 1.
\label{eq:nemql}
\end{equation}
which is remarkably in good agreement with 
the large lepton mixing recently observed\cite{SK}.

This relation can be obtained from more general framework 
using a kind of $SO(10)$ GUT and 
may be called the second q-l relation, similar to the first 
q-l relation, i.e., the celebrated bottom-tau
mass ratio of  $SU(5)$\cite{btau}. 
It is interesting that this relation can be 
most easily obtained from the twisted $E_6)$  
model. 
We would like to remark that our $E_6$ twisted model 
can also explain why the bottom quark mass is 
smaller  by almost $\lambda^2$ than  that of the top quark. 
This also comes from  the Higgs mixing factor 
$\sin \theta$ in $M_d$. 

To conclude, we have found that the twisted $E_6$ model can 
explain the up-down mystery ($m_t\gg m_b$), as well as 
the down-lepton mystery ($\theta_{\mu\tau}\gg \theta_{cb}$). 
It is well known that  $E_6$ gauge symmetry is naturally obtained 
from the 10 dimensional $E_8\times E_8$ heterotic string theory 
by the Calabi-Yau compactification into 4 dimensions. 
Our results are interesting and encouraging and may open the door for finding 
out more fundamental stringy GUTs including gravity. 
\vskip 2mm
This report is based on the works in collaboration with 
 T.~Kugo, K.~Yoshioka.  I would like to thank to T.~Yanagida, 
Y.~Nomura, M.~Yamaguchi and many other members for their 
stimulating discussions. 
This work has been done during the Summer Institute 98 and 99
held at Yamanashi, Japan organized by T. Kugo and T. Eguchi. 
I am supported in part by the Grants-in-Aid for Scientific Research  No.~9161 
from the Ministry of Education, Science, Sports and Culture, Japan.


\end{document}